\documentclass[%
twocolumn,
superscriptaddress,
 amsmath,amssymb,
 aps,
pra,
floatfix
]{revtex4}

\usepackage{color}

\usepackage{subfigure}

\usepackage{graphicx}% Include figure files
\usepackage{bm}
\usepackage{amsmath}
\usepackage{amssymb}

\usepackage{hyperref}

\usepackage[normalem]{ulem}

\begin{document}

\title{Machine learning with sub-diffraction resolution in the photon-counting regime}

\author{Giuseppe Buonaiuto}
\affiliation{Institute for High Performance Computing and Networking (ICAR), National Research Council (CNR), 
80131 Naples, Italy}
\author{Cosmo Lupo}
\affiliation{Dipartimento Interateneo di Fisica, Politecnico \& Universit\`a di Bari, 70126, Bari, Italy}
\affiliation{INFN, Sezione di Bari, 70126 Bari, Italy}

\date{\today}

\begin{abstract}
The resolution of optical imaging is classically limited by the width of the point-spread function, which in turn is determined by the Rayleigh length.
Recently, spatial-mode demultiplexing (SPADE) has been proposed as a method to achieve sub-Rayleigh estimation and discrimination of natural, incoherent sources.
Here we show that SPADE 
yields sub-diffraction resolution
in the broader context of 
image classification.
To achieve this goal, we outline a hybrid machine learning algorithm for image classification 
that 
includes a physical part and a computational part. 
The physical part implements a physical pre-processing of the optical field that cannot be simulated without essentially reducing the signal-to-noise ratio.
In detail, a spatial-mode demultiplexer is used to sort the transverse field, followed by mode-wise photon detection. 
In the computational part, the collected data are fed into an artificial neural network for training and classification. 
As a case study, we classify images from the MNIST dataset after severe blurring due to diffraction. Our numerical experiments demonstrate the ability to classify highly blurred images that would be otherwise indistinguishable by direct imaging without the physical pre-processing of the optical field.
\end{abstract}

\maketitle

%--

\section{Introduction}

Improving the resolution of optical imaging is a long-standing goal with a broad impact on science and technology, from astronomical observations to biomedical applications.
New schemes and methodologies are most commonly benchmarked against the well-known Rayleigh resolution criterion, which heuristically states that it is hard to resolve any detail smaller than the width of the point-spread function (PSF). The latter is of the order or the Rayleigh length $\mathrm{x_R} = \lambda D/R$, where $\lambda$ is the wavelength, $R$ the radius of the pupil of the optical system, and $D$ the distance to the object~\cite{Rayleigh1879}.

Contrary to common belief, the Rayleigh criterion does not represent by any means the fundamental limit to optical resolution.
In fact, this criterion only applies to direct imaging (DI), i.e.~pixel-by-pixel measurements of the field intensity.
In general, information is also carried by the phase of the quantum electromagnetic field, and can be extracted by an optical quantum computer, suitably programmed for the task of performing imaging with maximum resolution.
While a universal optical computer may require high-order nonlinear interaction, it has been proven that linear interferometry is optimal for such a task~\cite{TsangPRX,PhysRevLett.117.190801,PhysRevLett.117.190802,PhysRevLett.124.080503}. 
In particular, spatial-mode demultiplexing (SPADE) can achieve optimal performance in a variety of setting~\cite{HKrovi,lu2018quantum,Modern,GracePRL,PhysRevLett.127.130502,Schlichtholz:24,
PhysRevA.96.062107,Prasad,astro22}.

%%%

The primary goal of this work is to address the problem of image classification, in the sub-Rayleigh regime and in the limit of ultra-weak incoherent sources, using data collected from a finite number of photon detection events.
We focus on detection schemes that are derived from a particular interferometric scheme, dubbed Hermite-Gaussian (HG) SPADE~\cite{TsangPRX}, where the optical field is sorted in its transverse components along the HG modes~\cite{paur2016achieving,Paur:18,Zhou2019,Lvovsky,Zhou:23,Frank:23,Tham2017,Salit:20,Boucher2020,Santamaria22,Treps23,Amato23}. 
Here we consider both HG SPADE and its extensions where linear combinations of the HG modes are considered.

Until now, most applications of SPADE have been limited to the problems of parameter estimation and binary hypothesis testing.
In the latter (see e.g.~\cite{GracePRL,lu2018quantum,PhysRevLett.127.130502,astro22}) 
two quantum states,
$\rho_0$ and $\rho_1$, 
are given, 
and the goal is to find an optimal measurement to discriminate them
($M$-ary discrimination was also considered~\cite{GracePRL}).
Here we address a different problem.
We consider a large database of quantum states $\rho_{j,k}$, where $j=0,1$ labels two classes of states --- we also study the multiclass scenario --- and $k=1,\dots, K$ indicates the elements in each class.
Our goal is to find an optimal measurement and post-processing algorithm to associate quantum states to the different classes. 
While work in this direction has been reported with bright signals and coherent detection~\cite{Lvovsky,Frank:23}, to the best of our knowledge our contribution is the first attempt to address image classification in the photon-counting regime.

We remark that the two problems (hypothesis testing and classification) are related but different.
It is easy to show that they are equivalent only in the single-shot scenario (where only one photon is detected) and otherwise inequivalent for $N>1$ photon detections~\footnote{ 
Consider the problem of $M$-ary classification of quantum states, 
$\rho_{j,k}$, where $j=1,\dots, M$ labels the classes, 
and $k=1,\dots,K$ labels the elements in each class.
Consider a measurement modeled by the POVM elements $\Lambda_m$, for $m=1,\dots,M$.
If only one copy of the unknown state is available, then the average probability of successful classification is
\begin{align}
P_s = \frac{1}{MK} \sum_k \sum_j \mathrm{Tr}\left( \Lambda_j \rho_{j,k} \right)
= \frac{1}{M} \sum_j \mathrm{Tr}\left( \Lambda_j \rho_j \right) \, ,
\end{align}
where 
\begin{align}
 \rho_j = \frac{1}{K} \sum_k \rho_{j,k}   
\end{align}
denote the ensemble averages within each class.
This shows that, in the single-copy scenario, the classification problem is equivalent to the problem of hypothesis testing defined on the average states $\rho_j$.
In the multi-copy scenario, this equivalence is lost. 
In the multi-copy scenario we are allowed to measure $N$ copies of the unknown state using a POVM with elements $\Lambda_m^{(N)}$.
The probability of successful classification is
\begin{align}
P_s^{(N)} = \frac{1}{M K} \sum_k \sum_j \mathrm{Tr} \left( \Lambda_j^{(N)} \rho_{j,k}^{\otimes N} \right)
\end{align}
This is not the same as the probability of successful discrimination of the averages states since
\begin{align}
\frac{1}{K} \sum_k \rho_{j,k}^{\otimes N}
\neq \rho_j^{\otimes N} \, .
\end{align}
}.

As a case study we consider the MNIST (Modified National Institute of Standards and Technology) database of handwritten digits.
Our application may be viewed as a hybrid machine learning algorithm for image classification.
The physical part of our scheme is the application of SPADE, which implements a pre-processing of the information encoded in the unknown quantum state of a single photon.
The computational part is a machine learning algorithm, which takes the measurement outcomes as inputs data for training and classification

Here we simulate the physical part of the scheme, including the source emission, light propagation, and detection by SPADE, or DI for comparison, assuming imaging through a diffraction-limited optical system (e.g., a microscope or telescope) in the regime of ultra-weak signals. 
Samples of $N$ photon detection events are obtained through Monte Carlo simulation.
The simulated data are then fed into a machine learning algorithm.
Our numerical experiments show that HG SPADE (or its modifications) combined with machine learning allows us to classify images with accuracy much above what can be achieved by DI.

%---

The manuscript develops as follows.
In Section \ref{sec:model} we introduce the physical model and the relevant assumptions of our work.
In Section \ref{sec:HG} we present the expansion of a generic weak incoherent state in the HG basis.
In Section \ref{sec:ne} we illustrate our numerical experiments, including results methodology, for both binary and multiclass classification.
Finally, in Section \ref{sec:conc} we present our conclusions and outline potential future directions of research.

\section{The physical model}\label{sec:model}

Consider an extended, nearly planar source of incoherent, quasi-monochromatic light at wavelength $\lambda$. 
Such an object is characterised by its source intensity distribution.
For example, for our numerical experiments we choose objects from the MNIST dataset: grey-scale masks digits from $0$ to $9$, of $28 \times 28$ pixels, with intensity $I_{xy}$, where $x,y$ is the pixel location in the object plane. 
%Our physical model is schematically shown in Fig.~\ref{fig:imaging}.

We are interested in highly attenuated sources, emitting much less than one photon per pulse in average. In this regime, 
the light emitted by the object can be expressed as an incoherent sum of the vacuum and single-photon states, with density matrix
\begin{align}\label{state0}
\rho_0 = (1-\varepsilon) |0\rangle \langle 0| + \varepsilon \sum_{xy}  
I_{xy} |1_{xy}\rangle \langle 1_{xy}| \, ,
\end{align}
where $\varepsilon \ll 1$ is the probability that a photon is emitted,
$I_{xy}$ is the normalised intensity distribution,
$|0\rangle$ is the vacuum state, 
and
$|1_{xy}\rangle = a_{xy}^\dag|0\rangle$ is the state of a single photon emitted from the pixel in $x,y$, 
with
$a_{xy}^\dag$ being the corresponding bosonic creation operator.
Terms with two or more photons are negligible in the limit of ultra-weak signals.
For simplicity we consider a scalar field as polarisation-related effects are out of the scope of this work.

The object is observed using an optical imaging system, which in the far field and within the paraxial approximation is characterised by its PSF, denoted $T$~\cite{goodman2008introduction}. 
The PSF is the image of a point-like source. 
In particular, $T(x'-Mx,y'-My)$ is the amplitude of the field at position $x',y'$ in the image plane given that it was emitted at position $x,y$ in the object plane, where $M$ is the magnification factor (to simplify the notation, from now on we put $M=1$).
The particular form of the PSF depends on the details of the optical system. As a concrete example, here we consider a Gaussian PSF,
\begin{align}\label{PSF}
T(x'-x,y'-y) = \mathcal{N}^{1/2} e^{ - \frac{(x'-x)^2 + (y'-y)^2 }{4 \sigma^2} } \, ,
\end{align}
where $\mathcal{N}^{1/2}$ is the normalisation factor.

The width of the PSF quantifies the effects of diffraction in the optical system, and is of the order of the Rayleigh length, $\sigma \simeq \mathrm{x_R}$.
The Gaussian PSF can be used to approximate the PSF arising from a circular pupil, especially if $\sigma$ is much larger than the size of the source~\cite{Santamaria22}.

The optical imaging system maps the single-photon state $|1_{xy}\rangle$ into
\begin{align}
%|1_{x,y}\rangle  
%\to 
|T_{xy}\rangle := \sum_{x',y'} T(x-x',y-y')|1_{x'y'}\rangle 
\, .
\end{align}
Therefore, the source state in Eq.~(\ref{state0}) is mapped into the following state of the field in the image plane:
\begin{align}\label{state1}
\rho = (1-\varepsilon) |0\rangle \langle 0| 
+ \varepsilon \sum_{xy}
I_{xy} |T_{xy}\rangle \langle T_{xy}| \, .
\end{align}

We remark that, given two point-sources at position $(x,y)$ and $(u,v)$ in the object plane, we have $\langle T_{xy} |T_{uv}\rangle \neq 0$ unless the transverse separation between the point-sources is much larger than the width of the PSF. This means that the two sources cannot be perfectly resolved.

%---

\begin{figure}[t!]
\centering
\includegraphics[width=\columnwidth]{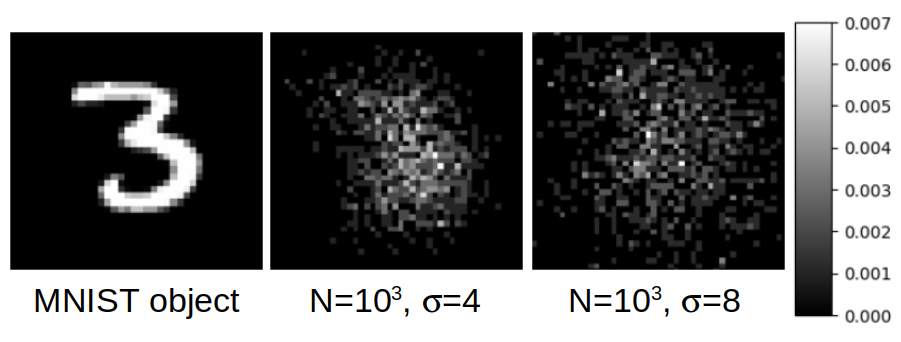}
\caption{On the left: an example of MNIST object. 
On the right: the relative frequencies of photon detection from DI, obtained through Monte Carlo simulation with $N=10^3$ photon detection events, for $\sigma=4,8$. 
The PSF width $\sigma$ is given in number of pixels. The color bar indicates the range of empirical relative frequencies.} 
\label{fig:imagesample}
\end{figure}

%---

\begin{figure*}[t!]
%\begin{figure}[t!]
\centering
\includegraphics[width=0.8\textwidth]{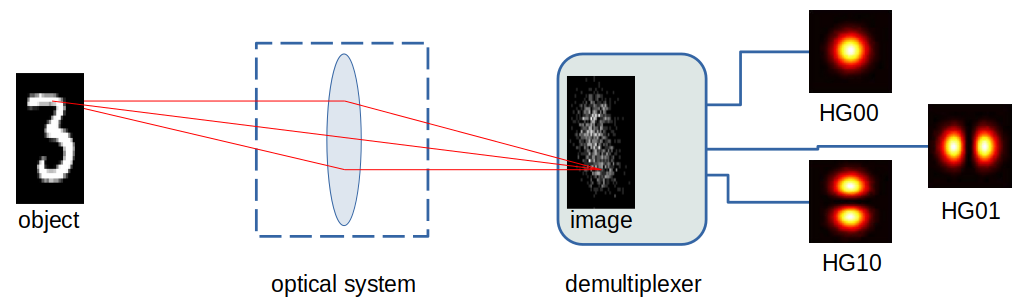}
\caption{The left part of the figure shows a schematic representation of image formation through an optical imaging system in the far field, paraxial approximation, and in the sub-Rayleigh regime.
The image is pixeleted to convey the fact that we consider only a finite number of photon detection events.
The right part of the figure schematically represents a measurement of the field by HG SPADE.
HG SPADE is a measurement strategy where the focused is processed by a spatial demultiplexer that sorts the transverse field in its Hermite-Gaussian component. 
Finally, each component is independently measured by photon counting. 
The figure shows only the lower modes $\text{HG}_{00}$, $\text{HG}_{01}$, $\text{HG}_{10}$.
In our work we also consider linear combinations of the HG states as well as higher modes.
} 
\label{fig:imaging}
\end{figure*}

%---

\subsection{Direct imaging}\label{subsec:DI}

Equation (\ref{state1}) represents the state of the field in the image plane. This is the state that is measured to extract information about the object.
DI consists of pixel-by-pixel photon detection in the image plane.
Conditioned on detecting a photon,
this yields measurement outcomes sampled from the probability distribution
\begin{align}
p_{x'y'}^\text{DI}
& = \sum_{xy} 
I_{xy} |\langle 1_{x'y'} |T_{xy}\rangle|^2 \\
& = \sum_{xy}  
I_{xy} | T(x-x',y-y') |^2 \, .
\end{align}

This probability density is the convolution of the object intensity distribution with the squared PSF.
For Gaussian PSF, this is a Gaussian convolution:
\begin{align}\label{pDI}
    p_{x'y'}^{\text{DI}} = \mathcal{N} 
    \sum_{xy} I_{xy}  \, e^{- \frac{(x'-x)^2+(y'-y)^2}{2\sigma^2}} \, .
\end{align}

Finally, we consider a finite number $N$ of photons detected. 
Therefore, the quantity of interest for us is the relative frequency, denoted $f_{x'y'}^{\text{DI}}(N)$, which is obtained by sampling (with replacement) from the probability distribution in (\ref{pDI}).

As an example, Fig.~\ref{fig:imagesample} shows an image in the MNIST database and the relative frequencies $f_{x'y'}^{\text{DI}}(N)$ obtained from the Monte Carlo simulation of $N = 10^3$ photon detection events, in correspondence of PSF width $\sigma = 4$ and $\sigma = 8$ (in units of number of pixels).

%---

\subsection{Spatial-mode demultiplexing}\label{subsec:SPADE}

Optics offers to the experimenter much more than DI. The optical field at the image plane can in fact be measured in other ways.
Among all possible detection strategies, a special role is played by SPADE, which is implemented through a linear, passive interferometer --- a schematic representation is shown in Fig.~\ref{fig:imaging}.
This implies that the quantum state of light in Eq.~(\ref{state1}) is processed without changing the photon number
(though in the presence of loss there is a non-zero probability that the photon is lost).

Consider a basis $\{ \Psi_{uv} \}_{u,v=0,1,\dots}$ of orthogonal functions in the image plane. They define a set of normal modes for the transverse field. In particular, consider the corresponding single-photon states, denoted as $|\Psi_{uv}\rangle$, that is,
\begin{align}
|\Psi_{uv} \rangle := \sum_{x,y} \Psi_{uv}(x,y)|1_{xy}\rangle 
\, .
\end{align}
The field can be decomposed along the $n^2$ lower modes by means of an interferometer with $n \times n$ output ports. 
A notable example is provided by HG modes, where
\begin{align}
\Psi_{uv}(x,y) \equiv \text{HG}_{uv}(x,y)
= \phi_u(x) \phi_v(y) 
\, ,
\end{align}
with
\begin{align}
\phi_u(x) =  
\frac{\mathcal{N}^{1/2}}{\sqrt{2^u u!}} \,
e^{ - \frac{ x^2 }{4 \sigma^2} }
H_u\left( \frac{x}{\sqrt{2}\sigma} \right) 
\, ,
\end{align}
and $H_u$ denote the Hermite polynomials.
Note that the width $\sigma$ matches that of the Gaussian PSF.

Let us denote as $b_{uv}^\dag$ the operator that creates a photon in the output port $(u,v)$ of the interferometer, for $u,v=1,\dots,n$.
Such an interferometer, when applied to single-photon states, is represented by the unitary operator
\begin{align}
    W = \sum_{uv} 
   |1_{uv}\rangle \langle \Psi_{uv}| \, ,
\end{align}
where $ |1_{uv}\rangle = b_{uv}^\dag |0\rangle$ is the state of a single photon in the $u,v$ output port.
By mode-wise photon detection on the output ports, we obtain the probability that the photon is measured in mode $\Psi_{uv}$: 
\begin{align} 
p_{uv}^\text{SPADE}
& = \sum_{xy}
I_{xy} |\langle 1_{uv} | W |T_{xy}\rangle|^2 \\
& = \frac{1}{\varepsilon } \langle \Psi_{uv} | \rho | \Psi_{uv} \rangle \, ,
\label{pSPADE}
\end{align}
with $\rho$ as in Eq.~(\ref{state1}), and the factor $1/\varepsilon$ comes from conditioning on photon detection.
A device that demultiplexes the optical field in the lower-order HG modes is schematically represented in Fig.~\ref{fig:imaging}.

Finally, for $N$ independent photon detection events, we need to consider the relative frequency $f_{uv}^{\text{SPADE}}(N)$, obtained by sampling with replacement from the probability distribution $p_{uv}^\text{SPADE}$.

%---

\section{Expansion in the HG basis}
\label{sec:HG}

In our numerical experiment we will make use of the an expansion of a generic single-photon state in terms of the HG modes.

First consider the field generated by a point-like source. This is expressed by the displaced PSF,
\begin{align}
|T_{x,y}\rangle 
& = \sum_{x',y'} T(x-x',y-y')|1_{x',y'}\rangle \\
& = \mathcal{N}^{1/2} \sum_{x',y'}  e^{ - \frac{(x-x')^2 + (y-y')^2 }{4 \sigma^2} } |1_{x',y'}\rangle \, ,
\end{align}
where again we are assuming a Gaussian PSF.
Expanding this vector in the HG basis we obtain
\begin{align}
|T_{x,y}\rangle 
& = e^{-\frac{|x|^2+|y|^2}{8 \sigma^2}} 
\sum_{m,n} \frac{1}{\sqrt{m!n!}} \frac{x^m y^n}{(2\sigma)^{m+n}} |\text{HG}_{mn}\rangle
\, .
\end{align}
From this we obtain the density matrix in Eq.~(\ref{state1}) conditioned on having at least one photon (i.e., we are neglecting the vacuum contribution)
\begin{align}
\tilde\rho 
& = \sum_{x,y} I_{x,y} |T_{x,y}\rangle \langle T_{x,y}| \\
%
%& = \sum_{x,y} I_{x,y} e^{-\frac{|x|^2+|y|^2}{4 \sigma^2}} \sum_{m,n} \frac{1}{\sqrt{m!n!}} \frac{x^m y^n}{(2\sigma)^{m+n}} |\text{HG}_{mn}\rangle \\
%& \phantom{=}~ \times \sum_{m',n'} \frac{1}{\sqrt{m'!n'!}} \frac{x^{m'} y^{n'}}{(2\sigma)^{m'+n'}} \langle \text{HG}_{m'n'} | \\
%
& = \sum_{x,y} I_{x,y} e^{-\frac{|x|^2+|y|^2}{4 \sigma^2}} \nonumber \\
& \phantom{=}~\times \sum_{mm'nn'} 
 \frac{x^{m+m'} y^{n+n'}}{(2\sigma)^{m+m'+n+n'}} 
 \frac{|\text{HG}_{mn}\rangle \langle \text{HG}_{m'n'} |}{\sqrt{m! m'! n! n'!}}
 \label{HGexpa}
\, .
\end{align}

%--

\section{Numerical experiments}\label{sec:ne}

In this Section we present a number of numerical experiments where HG SPADE, or a suitable modification thereof, is combined with machine learning algorithms into a hybrid approach to classify faint sources in the sub-Rayleigh regime. 
A working code for the preprocessing hereby performed can be found at~\footnote{\url{https://github.com/PeppeGoodhelp/Spade_Estimation/}}.
The dataset used for the experiments is the MNIST, handwritten digits, which is extracted from the \textit{torchvision} Python library. This is a widespread benchmark dataset in computer vision and machine learning. 

First we simulate the action of the optical imaging system, which focuses the field emitted from the object sources.
For experiments of binary classification, we consider restricted datasets of either $0$'s and $1$'s (in Section~\ref{subsec:01}) or $6$'s and $9$'s (in Section~\ref{subsec:69}). 
All classes of MNIST objects, i.e.~the digits from zero to nine, are considered in Sections~\ref{subsec:multi} and~\ref{subsec:69}.
Diffraction through the pupil of the optical system induces blurring. In our simulation we fix the value of the width of the PSF to $\sigma=9.5$ (measured in number of pixels), and scale the source object by a factor $f$: this induces a relative width of $\tilde \sigma = \sigma/f$.
From the focused optical field we can compute the probability of photon detection for both DI and HG SPADE, respectively using Eq.~(\ref{pDI}) and Eq.~(\ref{pSPADE}).
The statistics for given number of detected photon $N$ is generated by Monte Carlo simulation.

Finally, the datasets obtained in this way are directly ingested into a machine learning model trained for classification.
The models used in the following for performing the classification are standard ones in the machine learning literature: Random Forest \cite{breiman2001} and Fully Connected Neural Network (FCNN) \cite{Goodfellow}. 
RF is an ensemble learning method that proves to be quite effective both in regression and in classification. Furthermore, it is relativity light, hence, given the size of the HG SPADE vector, it is expected to capture the fundamental features useful for the classification. 
FCNN on the contrary is a deep learning method, particularly used for classification, that extracts features from the input sample via a series of connected hidden layers with non-linear activation functions. It is one of the standard models for automated classification, particularly useful in computer vision.

%--

\begin{figure}[t!]
\centering
\includegraphics[width=0.8\columnwidth]{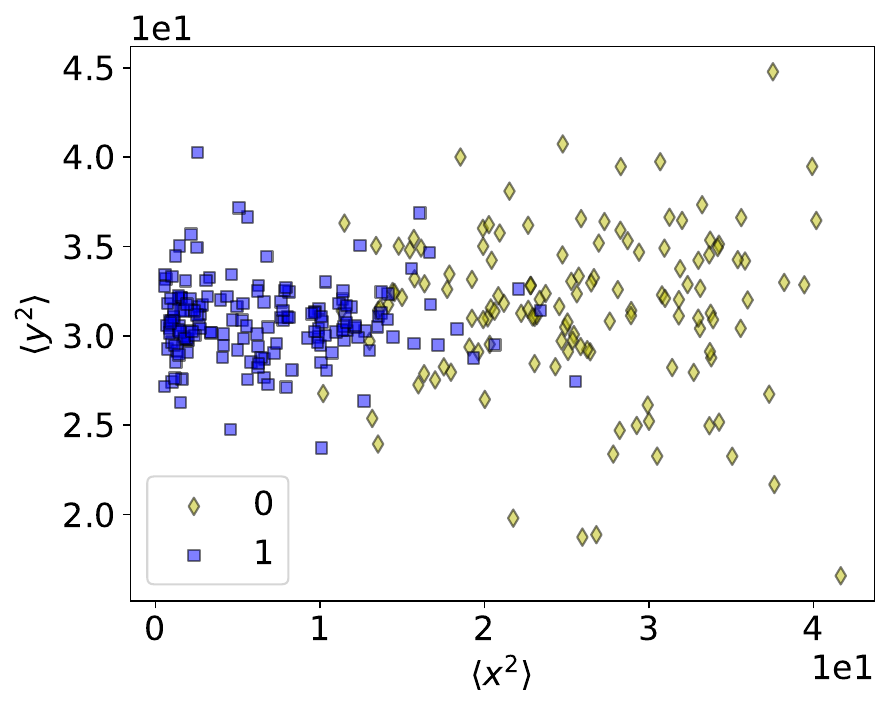}\\
\includegraphics[width=0.8\columnwidth]{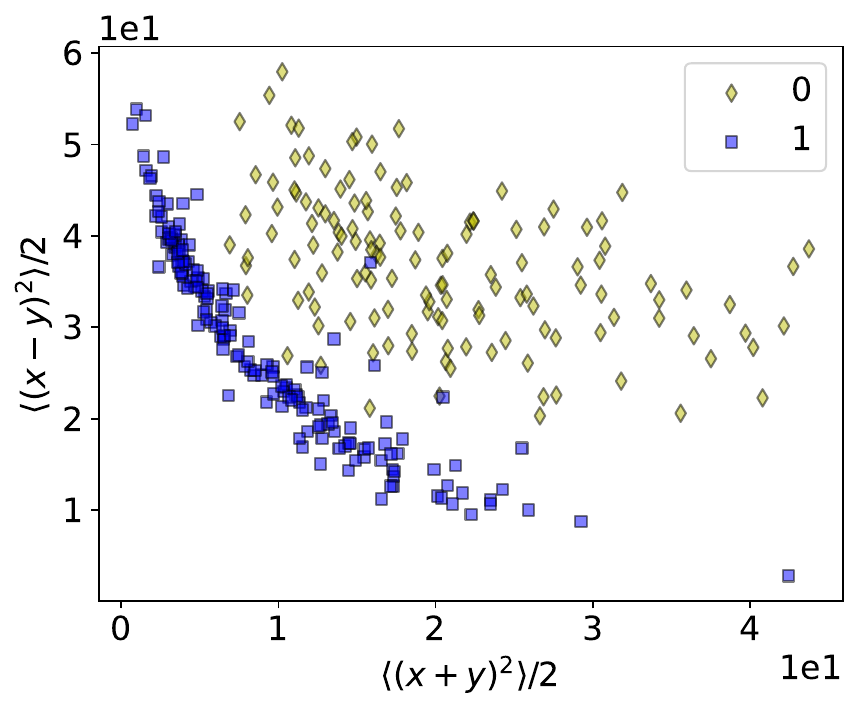}
\caption{Scatter plot of the second moments of a sample of $300$ objects from the MNIST classes $0$ and $1$. 
Top: the moments are computed in the Cartesian basis, according to Eqs.~(\ref{C1})-(\ref{C2}). Bottom: moments in the diagonal basis, computed as in Eqs.~(\ref{D1})-(\ref{D2}).
} 
\label{fig:moments2}
\end{figure}

%--

\subsection{Binary classification, of $0$'s vs $1$'s}\label{subsec:01}

The first problem we consider is the binary classification of MNIST objects from the classes of $0$'s and $1$'s. 
Elements belonging to these two classes are well separated in terms of the second moments of their intensity. 
Figure~\ref{fig:moments2} shows the distribution of second moments for a sample of $300$ objects from the database.
the figure shows the moments in both the Cartesian basis,
\begin{align}
\langle x^2 \rangle & = \sum_{x,y} I_{x,y} x^2 \, , \label{C1} \\
\langle y^2 \rangle & = \sum_{x,y} I_{x,y} y^2 \, , \label{C2}
\end{align}
and in the diagonal basis (obtained by rotation of $45^\circ$ degrees),
\begin{align}
\frac{1}{2}\langle (x+y)^2 \rangle & = \frac{1}{2} \sum_{x,y} I_{x,y} (x+y)^2 \, , \label{D1} \\
\frac{1}{2}\langle (x-y)^2 \rangle & = \frac{1}{2} \sum_{x,y} I_{x,y} (x-y)^2 \, . \label{D2}
\end{align}

Note that the MNIST objects are centered in such a way that we can put 
$\langle x \rangle = \langle y \rangle = 0$.
In our numerical experiment we focus on extracting information about the moments in the diagonal basis, as they appear to better separate the two classes.

%--

\subsubsection{SPADE:}

From Eq.~(\ref{HGexpa}) we see that information about the second moments in Eqs.~(\ref{D1})-(\ref{D2}) can be obtained through a measurement in the basis of rotated lower HG modes:
\begin{align}
\left\{
|\text{HG}_{00}\rangle ,
\frac{ |\text{HG}_{10}\rangle + |\text{HG}_{01}\rangle }{\sqrt{2}},
\frac{ |\text{HG}_{10}\rangle - |\text{HG}_{01}\rangle }{\sqrt{2}}
\right\}
\, .
\end{align}

The probability of detecting a photon in these modes can be computed directly (up to normalisation) from Eq.~(\ref{HGexpa}):
\begin{align}
|\text{HG}_{00}\rangle 
& \to  
\sum_{x,y} I_{x,y} e^{-\frac{|x|^2+|y|^2}{4 \sigma^2}} 
= p_{00}^\text{SPADE} 
\, , 
\label{pHG00}\\
%%%
\frac{ |\text{HG}_{10}\rangle \pm |\text{HG}_{01}\rangle }{\sqrt{2}}
& \to  
\frac{1}{8 \sigma^2} 
\sum_{x,y} I_{x,y}    
 ( x \pm y )^2 
 e^{-\frac{|x|^2+|y|^2}{4 \sigma^2}} 
 \nonumber \\
& \phantom{\to}= p_{\pm}^\text{SPADE} 
 \, . 
 \label{pHGpm}
 %\\
%%%
%\frac{ |\text{HG}_{10}\rangle + |\text{HG}_{01}\rangle }{\sqrt{2}}
%& \to  
%\frac{1}{8 \sigma^2} \sum_{x,y} I_{x,y} ( x-y )^2  e^{-\frac{|x|^2+|y|^2}{4 \sigma^2}}
%= p_{-}^\text{SPADE} & = 
%\, .
\end{align}

%--

\begin{figure*}
    \centering
    \includegraphics[width=0.7\linewidth]{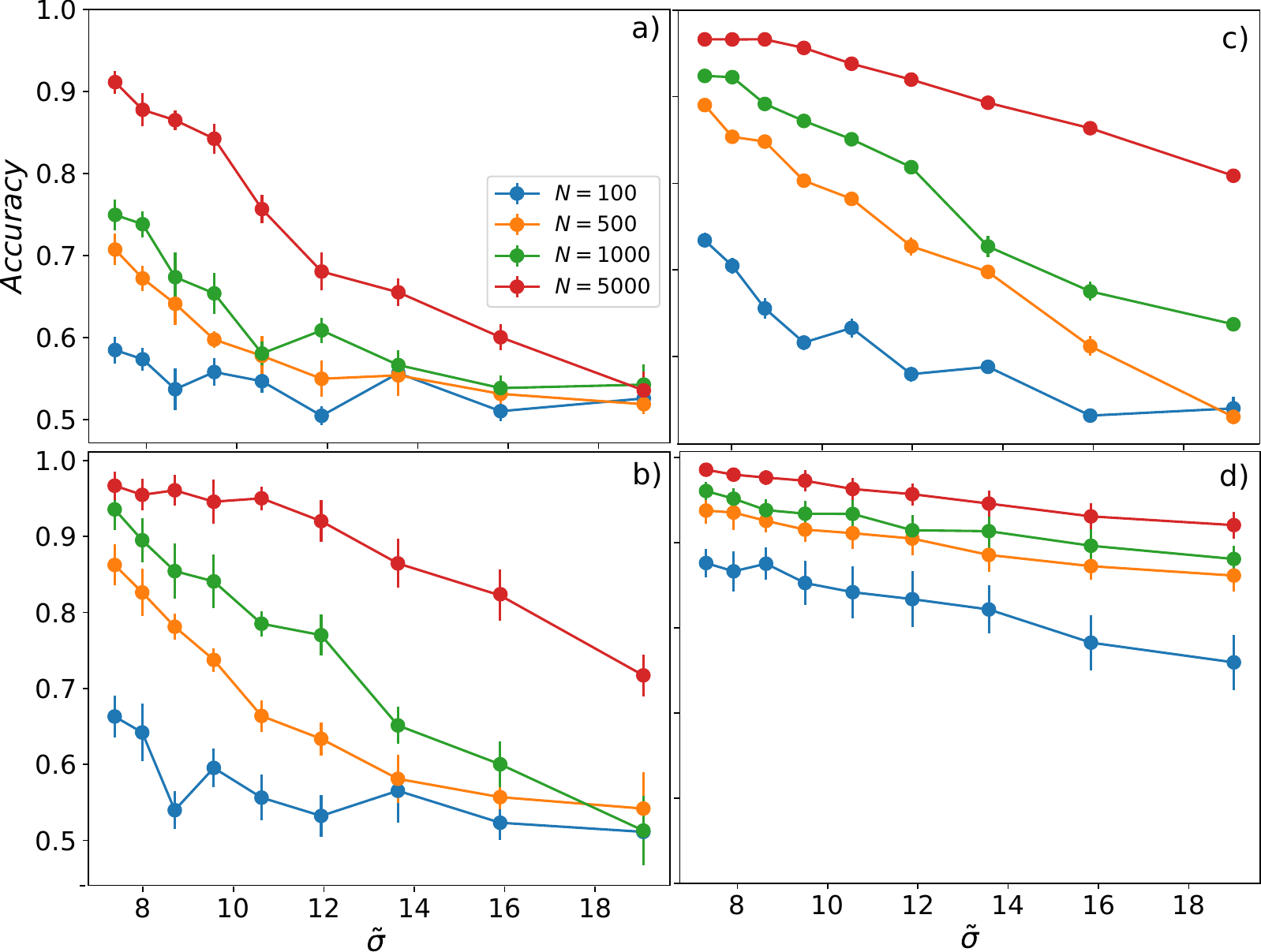}
    \caption{
    Accuracy in the binary classification of $0$'s and $1$'s in the MNIST dataset.
    The four panels are obtained from different data and using different machine learning algorithms:
    (a) simulated DI data processed with RF;
    (b) simulated DI data processed with FCNN;
    (c) second moments computed from simulated DI data, processed with RF;
    (d) simulated SPADE data, evaluated in a rotated basis as in Eq.~(\ref{pHGpm}), processed with RF.
    The results are evaluated for different effective blurring factors $\tilde \sigma = 9.5/f$ and considering various values of the number of photon detection events, $N$ (from bottom to topo, $N=100, 500, 1000, 5000$). The error bars are calculated via stratified $k$-fold \cite{stratifiedk}, using $m=10$ splits of the test dataset.
    } 
    \label{fig:DI+SPADE}
\end{figure*}

For any given integer value for the photon number $N$, we have implemented a Monte Carlo simulation to generate the relative frequencies
(see Section~\ref{subsec:SPADE}) by sampling (with replacement) from the above probability mass distributions.
We have tested a RF using these relative frequencies as feature vectors.
With a sample of $11867$ images from the database, and using $70\%$ of them for training and $30\%$ for validation.
The results are displayed in Fig.~\ref{fig:DI+SPADE}(d), which shows the accuracy vs the PSF width $\tilde \sigma$, for different values of the photon number $N$.

Figure~\ref{fig:DI+SPADE}(d) shows the qualitative behaviours of the accuracy:
(1) the accuracy decreases with increasing $\tilde \sigma$, as the Gaussian noise induced by diffraction makes more difficult to discriminate the objects;
(2) the accuracy increases with increasing $N$, due to smaller statistical fluctuations.

%--

%

%--

\subsubsection{DI:}

To compare with SPADE, we have performed a similar analysis for DI. For any number $N$ of photon detection events, we have generated the relative frequencies $f_{x'y'}^{\text{DI}}(N)$ (see Section~\ref{subsec:DI}) by sampling (with replacement) from the DI probability distribution in Eq.~(\ref{pDI}).
We have tested the machine learning approach, with results displayed in Fig.~\ref{fig:DI+SPADE}, panels (a), (b) and (c).
Panel (a) shows the results obtained by using RF, and panel (b) is for a FCNN.
A direct comparison suggests that the FCNN performs much better than RF for this problem, as the hidden neurons are able to extract relevant features out of the original sample. We argue that, in particular, the FCNN is able to obtain the information most likely related to the moments of order two, which is crucial to classify correctly in the sub-Rayleigh regime.

Finally, also for a fair comparison with SPADE, we helped the algorithm by performing a pre-computation of the second moments. 
Starting from the relative frequencies $f_{x'y'}^{\text{DI}}(N)$, the second moments can be estimated as follows:
\begin{align}
\langle x^2 \rangle & \leftarrow \sum_{x',y'} f_{x'y'}^{\text{DI}}(N) {x'}^2 \, , \\
\langle y^2 \rangle & \leftarrow \sum_{x',y'} f_{x'y'}^{\text{DI}}(N) {y'}^2 \, ,
\end{align}
for the Cartesian basis, and 
\begin{align}
\frac{1}{2}\langle (x+y)^2 \rangle & \leftarrow \frac{1}{2} \sum_{x',y'} f_{x'y'}^{\text{DI}}(N) (x'+y')^2 \, , \label{D1DI} \\
\frac{1}{2}\langle (x'-y')^2 \rangle & \leftarrow \frac{1}{2} \sum_{x',y'} f_{x'y'}^{\text{DI}}(N) (x'-y')^2 \, ,
\label{D2DI}
\end{align}
for the diagonal one.
In particular, we have used the estimates of the second moments in diagonal basis and used them as feature vectors in a RF. 
The results are displayed in the bottom panel (c) of Fig.~\ref{fig:DI+SPADE}, showing an increase in the accuracy compared to the FCNN fed with the relative frequencies. Moreover, it can be noticed that the fluctuations for each $\tilde{\sigma}$ are more contained compared to the previous cases: this is due to the fact that the second moments, which now are structured within the feature vector, are independent of $\tilde{\sigma}$ and hence they stabilise the computation.

In conclusion, DI supported by either RF and FCNN yields poor performances when compared against SPADE supported by RF. 
The accuracy of DI increases when the second moments are pre-computed from the DI relative frequencies and fed into a RF. 
This latter approach mimics the pre-processing implemented by SPADE, though in a 
computational rather than 
physical way. However, it also 
performs poorly in comparison with SPADE, especially for low photon number and severe blurring.
This is due to the fact that computing the moments from the data obtained by DI yields a signal-to-noise ratio smaller than a direct measurement of SPADE~\cite{Santamaria22}.

%---

\begin{figure}[t!]
\centering
\includegraphics[width=1.1\columnwidth]{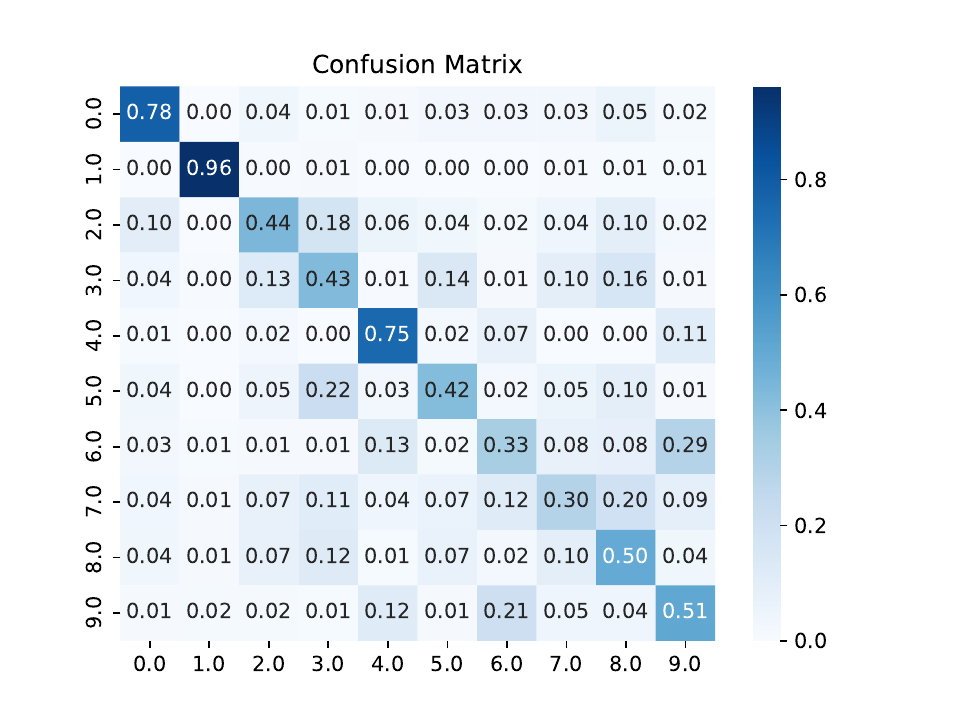}
\caption{
Confusion matrix of the output of the RF for multi-class classification, performed on SPADE vectors. 
In particular the figure refers to the case with an effective blurring factor $\tilde{\sigma}=10$ and with number of photon detection events $N=5000$. 
In this numerical experiment, half of the photons are measured in the diagonal basis [Eq.~(\ref{pHGpm})] and the other half in the Cartesian basis [Eqs.~(\ref{HGp10})-(\ref{HGp01})].
} 
\label{fig:multiSPADE}
\end{figure}

% --

\subsection{Multiple classes}\label{subsec:multi}

Here we consider MNIST objects from all classes of digits from $0$ to $9$.
To test the accuracy of SPADE for this multi-class problem we considered features arising from a measurement of lower HG modes, in both the HG basis and its rotated version.

For the rotated HG basis, the probabilities are given in Eq.~(\ref{pHG00})-(\ref{pHGpm}).
For the un-rotated basis, photon detection is simulated in the lower-order HG modes:
\begin{align}
\left\{
|\text{HG}_{00}\rangle ,
|\text{HG}_{10} \rangle,
|\text{HG}_{01} \rangle
\right\}
\, ,
\end{align}
yielding the probabilities $p_{00}^\text{SPADE}$ in (\ref{pHG00}) and
\begin{align}
|\text{HG}_{10}\rangle 
& \to  
\frac{1}{4 \sigma^2} 
\sum_{x,y} I_{x,y}    
 x^2 
 e^{-\frac{|x|^2+|y|^2}{4 \sigma^2}} 
= p_{10}^\text{SPADE} 
 \, , 
 \label{HGp10} \\ 
%%%
|\text{HG}_{01}\rangle 
& \to  
\frac{1}{4 \sigma^2} 
\sum_{x,y} I_{x,y}    
 y^2 
 e^{-\frac{|x|^2+|y|^2}{4 \sigma^2}} 
= p_{01}^\text{SPADE} 
 \, .
  \label{HGp01}
 \end{align}

In our Monte Carlo simulations, we assume that upon $N$ photons detected, half are measured in the Cartesian basis and half in the diagonal basis.

The obtained confusion matrix is shown in Fig.~\ref{fig:multiSPADE} for $N=5000$ and $\tilde\sigma=10$, where each row is associated with the ground truth and the columns represent the guessed class.
The matrix shows mixed results. 
As expected, SPADE is able to discriminate very well between $0$'s and $1$'s. However, a lower accuracy is obtained for other digits. An example is given by the classes of $6$'s and $9$'s, which are essentially indistinguishable. %from their second moments.
This result is easily explained by noticing that, whereas the $0$'s and $1$'s are well separated by the second moments, this is not in general the case for the other digits.
This suggests that we may need to go beyond the second moments, i.e., include in our simulation and detection HG modes of higher degrees above $\text{HG}_{01}$ and $\text{HG}_{10}$.

% --

\subsection{Higher-order HG modes}
\label{subsec:69}

To extend the use of SPADE beyond second moments, we include in our simulations the events of photon detection in higher-order HG vectors, obtained from linear combinations of
$|\text{HG}_{20}\rangle$, 
$|\text{HG}_{02}\rangle$. 

\begin{figure}[t!]
\centering
\includegraphics[width=0.8\columnwidth]{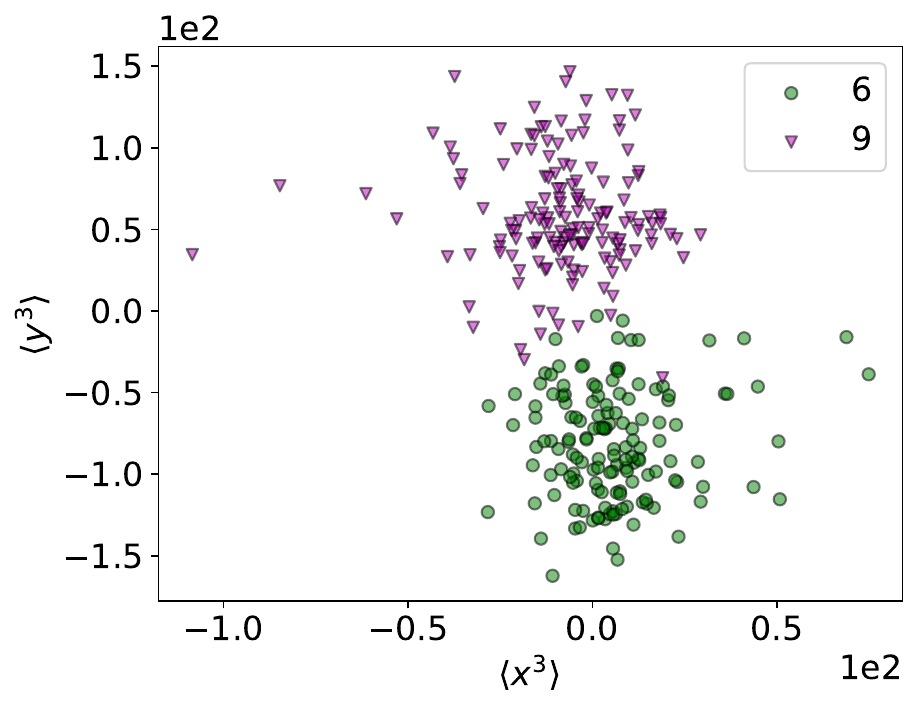}
\caption{
Scatter plot of the third moments of a sample of $300$ objects from the MNIST classes $6$ and $9$.} 
\label{fig:scheme_69}
\end{figure}

As a first concrete example, we focus on the problem of classifying MNIST objects belonging to the classes of $6$'s and $9$'s.
Objects from these classes have essentially identical distributions of the second moments, but
are well separated by the third moment $\langle y^3 \rangle$, as seen from Fig.~\ref{fig:scheme_69}.
We therefore consider measurements in the six vectors (this choice is not necessarily optimal, but provides a way to extract information about $\langle y^3 \rangle$):
\begin{align} \label{newspade}
\left\{
|\text{HG}_{00}\rangle ,    
\frac{ |\text{HG}_{01}\rangle \pm |\text{HG}_{02}\rangle}{\sqrt{2}} ,
|\text{HG}_{10} \rangle ,
|\text{HG}_{20}\rangle , 
|\text{HG}_{11}\rangle
\right\} \, .
\end{align}

Photon detection in these states is related to combination of moments up to the fourth order, including
\begin{align}
%& |\text{HG}_{00}\rangle \to 1 - \frac{\langle x^{2} \rangle + \langle y^{2} \rangle }{(2\sigma)^{2}} - \frac{ \langle x^{4} \rangle + \langle x^{2} y^{2} \rangle + \langle y^{4} \rangle }{(2\sigma)^{4}} \, , \\
%
& \frac{ |\text{HG}_{01}\rangle \pm |\text{HG}_{02}\rangle}{\sqrt{2}} \nonumber \\
& \to 
\frac{1}{2}
\sum_{x,y} I_{x,y} e^{-\frac{|x|^2+|y|^2}{4 \sigma^2}} 
\left(
    \frac{ y^{2} }{(2\sigma)^{2}} 
    + \frac{ y^{4} }{2(2\sigma)^{4}}
    \pm 
    \frac{ \sqrt{ 2 } y^{3}}{(2\sigma)^{3}} 
\right) \\
& \sim
    \frac{1}{2} \left(
    \frac{\langle y^{2} \rangle }{(2\sigma)^{2}} 
    + \frac{ \langle y^{4} \rangle }{2(2\sigma)^{4}}
    \pm 
    \frac{ \sqrt{ 2 } \langle y^{3} \rangle}{(2\sigma)^{3}} 
\right) \, .
%
%& \frac{ |\text{HG}_{10}\rangle \pm |\text{HG}_{20}\rangle}{\sqrt{2}} \to \frac{1}{2} \left( \frac{ \langle x^{2} \rangle }{(2\sigma)^{2}} + \frac{ \langle x^{4} \rangle }{2(2\sigma)^{4}} \pm \frac{ \sqrt{ 2 } \langle x^{3} \rangle }{(2\sigma)^{3}} \right) \, , \\
%%
%& |\text{HG}_{11}\rangle  \to \frac{ \langle x^{2} y^{2} \rangle}{(2\sigma)^{4}} \, .
\end{align}

A partial optimisation may be obtained by introducing an angle of rotation:
\begin{align}
& \sin{\varphi} |\text{HG}_{01}\rangle + \cos{\varphi} |\text{HG}_{02}\rangle
    \to \nonumber \\
    & \hspace{0.3cm} (\sin{\varphi})^2 \frac{ \langle y^{2} \rangle }{(2\sigma)^{2}} 
    + (\cos{\varphi})^2 \frac{ \langle y^{4} \rangle}{2(2\sigma)^{4}}
    + 
    \sqrt{2} \sin{\varphi}\cos{\varphi}\frac{ \langle y^{3} \rangle }{(2\sigma)^{3}} 
 \, , \\
& \cos{\varphi} |\text{HG}_{01}\rangle - \sin{\varphi} |\text{HG}_{02}\rangle
    \to \nonumber \\
    & \hspace{0.3cm} (\cos{\varphi})^2 \frac{ \langle y^{2} \rangle }{(2\sigma)^{2}} 
    + (\sin{\varphi})^2 \frac{ \langle y^{4} \rangle }{2(2\sigma)^{4}}
    - 
    \sqrt{2} \sin{\varphi} \cos{\varphi} \frac{\langle y^{3} \rangle }{(2\sigma)^{3}} 
 \, .
\end{align}

The angle $\varphi$ is optimised to maximise the signal-to-noise ratio, where the signal is represented by the moment of order three and the noise by the moments of order two and four. 
By putting 
\begin{align}
\sin{\varphi} & = \frac{2\sigma}{\sqrt{ (2\sigma)^2 + 2(2\sigma)^{4}} } \, , \\
\cos{\varphi} & = \frac{\sqrt{2}(2\sigma)^{2}}{\sqrt{ (2\sigma)^2 + 2(2\sigma)^{4}} } \, ,
\end{align}
we obtain
\begin{align}
\sin{\varphi} |\text{HG}_{01}\rangle + \cos{\varphi} |\text{HG}_{02}\rangle
    & \to 
%\nonumber \\
%& \hspace{0.1cm} 
\frac{ \langle y^{2} \rangle + \langle y^{4} \rangle + 2 \langle y^{3} \rangle}{(2\sigma)^2 + 2(2\sigma)^{4}} \, , \\
\cos{\varphi} |\text{HG}_{01}\rangle - \sin{\varphi} |\text{HG}_{02}\rangle
   & \to \nonumber \\    
& \hspace{0.1cm} \frac{ 2 
    (2\sigma)^{2} \langle y^{2} \rangle 
    + \frac{ \langle y^{4} \rangle}{2(2\sigma)^{2}}
    - 
    2 
    \langle y^{3} \rangle}{(2\sigma)^2 + 2(2\sigma)^{4}}
    \, .
\end{align}

Finally, to ensure normalisation, we also include in the simulation the probabilities of detecting a photon in the higher-order terms
$\text{HG}_{03}$, 
$\text{HG}_{12}$, 
$\text{HG}_{21}$,
$\text{HG}_{30}$.

\begin{figure}[t!]
\centering
\includegraphics[width=0.8\columnwidth]{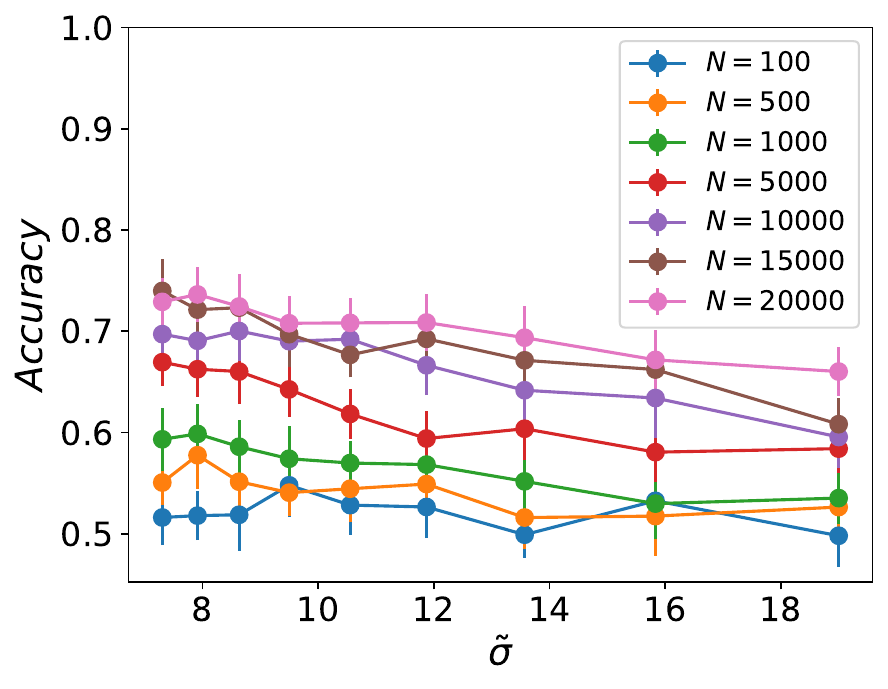}\\
\includegraphics[width=0.8\columnwidth]{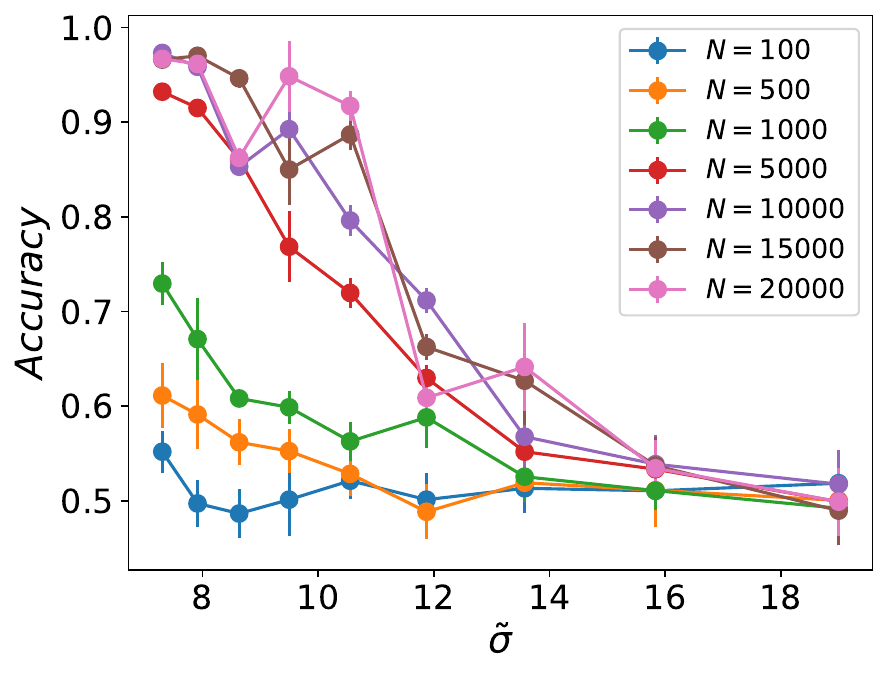}
\caption{
Accuracy vs blurring parameter $\tilde \sigma$ for the classification of $6$'s and $9$'s.
This is shown for several values of the number of detected photon, from bottom to top $N=100, 500, 1000, 5000, 10000, 15000, 20000$.
Top panel: data collected using the modified SPADE defined in Section~\ref{subsec:69}, and processed using a RF.
Bottom panel: data collected using DI, and processed using FCNN.
} 
\label{fig:69compa}
\end{figure}

The results of our numerical experiments are shown in Fig.~\ref{fig:69compa}.
We have applied this modified SPADE using the above features (with the optimised angle $\varphi$), with $N$ photon detection events, in a RF model.
The accuracy is lower than in the case of $0$'s vs $1$'s, and it requires many more detection events (higher values of $N$). 
This is because the third-order moments are suppressed by a factor proportional to $\tilde\sigma^{-3}$. 
With increasing $\tilde\sigma$, it is much less likely that a photon is detected in an HG mode of higher order, as most of them end up in the low-order HG modes.

We then compare with DI, powered by FCNN. For small values of $\tilde\sigma$, DI works better than SPADE, but for larger values of $\tilde\sigma$ SPADE has better performance in terms of both accuracy and stability.
Overall, this modified SPADE remains superior in the deep sub-Rayleigh regime (larger values of $\tilde\sigma$), but more photon detection events are needed to acquire a decent signal-to-noise ratio.
In conclusion, photons detected in the rotated HG modes are rare yet highly informative.

%---

%---

%---

\begin{figure}[t!]
\centering
\includegraphics[width=1.1\columnwidth]{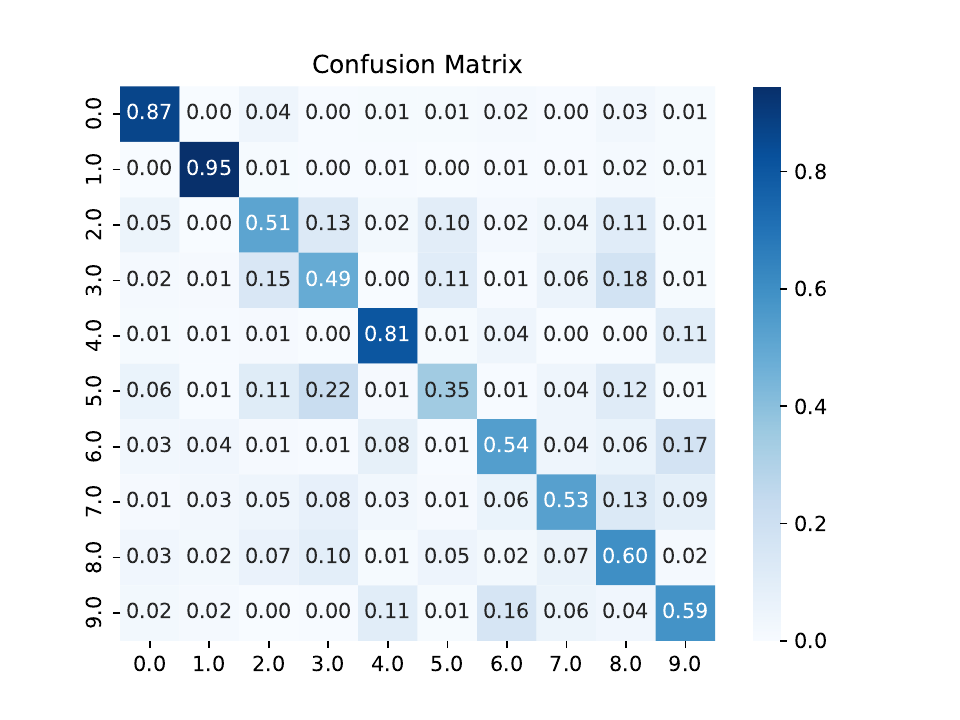}
\caption{
Confusion matrix of the output of the RF for multi-class classification, performed on modified SPADE in Eq.~(\ref{newspade}).
The figure refers to an effective blurring factor $\tilde{\sigma}=10$ and number of photon detection events $N=5000$. 
Compared with the confusion matrix in Fig.~(\ref{fig:multiSPADE}), this shows that including higher-order HG modes may substantially increase multi-class classification.
} 
\label{fig:multiSPADE3}
\end{figure}

Going beyond binary classification, we repeated the multi-class numerical experiment but this time using the modified SPADE. The results are shown in Fig.~(\ref{fig:multiSPADE3}). 
While in general the performance depends on which specific class is best characterised by the second of third moments, the confusion matrix in Fig.~(\ref{fig:multiSPADE3}) indicates that in most cases considering the higher-order HG modes yields higher accuracy in the multi-class classification problem.

%---

%---

\subsection{Methodology}

\subsubsection{MNIST Dataset:}

The full MNIST dataset, provided by the open access library \textit{torchvision}, consists of a collection of handwritten digits, with $70k$ samples. Each sample is a grayscale image of size $28 \times 28$ pixels, representing a single digit, with nominal pixel values ranging from $0$ to $255$. The dataset is widely considered a benchnmark for testing various machine learning and artificial intelligence algorithms, particularly those dedicated to image classification tasks. Even though it serves as a baseline, the MNIST dataset entails a diverse range of digit variations, including different writing styles, sizes and orientation. Hence it poses several challenges, such as variability in writing quality, noise and digit overlapping, making it an interesting test-bed for evaluating the effectiveness of the proposed methodologies. 
In this work, we make use of RF algorithm and FCNN, for evaluating the information content, and thus the performances in the classification, for the HG SPADE vectors and for the DI approach, as a function of various finite photon detection events and within a range of physical parameters.

For each model used in the experimental phase, the accuracy of the predictions is calculated by comparing them with the true labels in the test set:
\begin{equation}
    \text{Accuracy}=\frac{N_{c}}{N_{T}} \, , 
\end{equation}
where $N_{c}$ is the number of correct prediction, while $N_{T}$ is the total number of predictions.
The effectiveness of the SPADE technique is assessed based on the improvement in classification accuracy, in case of extreme blurring, compared to the results when the direct intensity features are used.

%--

\subsubsection{Data Preprocessing and Models Architecture:}

For binary classification, starting from the full MNIST dataset we consider the reduced problem of binary classification: out of the $10$ classes, corresponding to the first ten integer numbers, we select only two classes.
For the direct intensity case, the images are padded on an expanded grid of size $80 \times 80$ pixels: this operation is necessary for preventing the combination of high scaling factor and blurring to cut out of the original grid useful information, which might lead to biased results.  For the sake of computational time, which is greatly enhanced by the padded grid, we take a sub-set of the original dataset, reducing the size to $2000$ entries, balancing each of the two classes considered. Although this fact might, in principle, hinder the learning capabilities of the algorithm, both the trend of the loss function, and of the accuracy, are not effected, as they depend strongly upon the representation power of the feature vectors.  
For both DI and SPADE protocols we evaluate the performances for several manifolds of independent photon detection, ranging from low photo-detection frequency, i.e.~$N=100$, to higher values of $N$. The upper values of $N$ selected depend on whether or not using third order moments is necessary: in fact the probability of populating third order moments is low hence a greater collection of photons is necessary in order for these features to be meaningful for the classification. 
For SPADE, the feature vector is five-dimensional when moments up to the second order are considered, with entries extracted from the modes HG01, HG10, HG11, HG02, HG20. When third order moments are necessary, as in case of $6$ vs.~$9$ classification, the SPADE feature vector is ten-dimensional, with the additional entries extracted from the modes, HG21, HG12, HG30, HG03, HG02, HG20, HG11. 

The RF algorithm, both in the DI and SPADE case, is directly available from $Scikit-learn$, and has been trained with various hyperparameters, performing a standard grid-search, for optimising the performances. In particular, the number of estimators used in the computation of the RF is $200$. The FCNN instead has been constructed in \textit{Pytorch}, and consists of one layer of $256$ neurons, one layer of $128$ hidden neurons and a layer of $32$ hidden neurons. After each layer a ReLu activation function is in place to insert a non-linearity in the model. The final layer entails a Sigmoid function to estimate the probability of the sample of being in one of the two classes.

%--

\subsubsection{Training Procedure:}

The reduced MNIST dataset is divided into a training set and a validation set with an $70:30$ split. The training set is used to find the most suitable parameters for the model considered, while the validation set is employed to monitor the models performances. For FCNN, early stopping and dropout are in place, to prevent over-fitting. Finally, for the training procedure, the Adam optimiser has been used. 
The RF model minimises the loss function according to the~\textit{Gini} criterion \cite{Ginic}, which measures the probability of extracting a sample out of the two classes. The FCNN instead is trained to minimise the binary cross entropy between the target and the output of the model:
\begin{equation}
    L=\sum_{i=1}^{N} y_{n} \log(x_{n})+(1-y_{n})\log(1-x_{n}) \, ,
\end{equation} 
where $x_{n}$ is the target class and $y_{n}$ is the output class generated by the models. The hyperparameters selected for each model are listed in the result section.

%---

\section{Conclusions}\label{sec:conc}

We have explored a hybrid scheme for image classification, where the physical part is a hardware-based pre-processing of the optical field, followed by photon-detection.
The computational part is a machine learning algorithm, such as Random Forest (RF) or Fully Connected Neural Network (FCNN).
The hardware part is represented by a multi-mode interferometer that implements spatial-mode demultiplexing (SPADE) of the transverse field. In particular, we have focused on Hermite-Gaussian (HG) modes and their linear combinations.

Our work is the first attempt to exploit SPADE in the context of image classification and machine learning within the sub-Rayleigh and photon-counting regime.
With some exceptions, see e.g.~Refs.~\cite{Lvovsky,Frank:23}, most of the previous works were limited to problems of parameter estimation and hypothesis testing.

As a case study, we have applied this approach to classify objects from the MNIST database of handwritten digits from $0$ to $9$.
Our simulations and numerical experiments show that SPADE can classify objects that are otherwise indistinguishable using classical approaches based on DI.

To further improve the accuracy of our hybrid image classifier, one should optimise the choice of the demultiplexer that implements SPADE. Formally, this means optimise the choice of the unitary matrix that mathematically represents such a transformation.
That is, during the training phase, not only the parameters of the neural network are optimised, but also those of the unitary matrix~\cite{Spall}. 

Effects of experimental imperfections on the efficacy of SPADE, in particular cross-talk, have been assessed by a number of previous works, see for example Refs.~\cite{PhysRevLett.125.100501,Boucher2020,Santamaria22,Banaszek2020,PhysRevA.101.022323,PhysRevLett.126.120502}. 
In general noise and cross-talk decrease the efficacy of SPADE, but the advantage with respect to DI remains for mild imperfections. We expect the same to happen in our machine learning framework, but a detailed analysis is left for future works. 
We note that our approach may be combined with other schemes of quantum imaging and stellar interferometry, e.g.~\cite{app11146414,PhysRevLett.109.070503,PhysRevLett.123.070504,PhysRevLett.129.210502}.
Finally, our approach could be also extended to un-supervised learning, as well as to classes of objects of higher complexity than the MNIST digits.

%--

\section{Acknowledgements}\label{sec:ack}

This work has received funding from: 
the European Union --- Next Generation EU
through 
PNRR MUR project PE0000023-NQSTI
and
PRIN 2022 (CUP: D53D23002850006);
the Italian Space Agency (ASI, Agenzia Spaziale Italiana), project `Subdiffraction Quantum Imaging' (SQI) n.~2023-13-HH.0;
and INFN through the project `QUANTUM'.
We acknowledge discussions with Saverio Pascazio, Angelo Mariano, Francesco Pepe, and Seth Lloyd during the early stage of this project.

\section{Statements and Declarations}

Author contributions: G.B. performed the numerical experiments. 
C.L. designed the research and wrote the manuscript with the support of G.B.
All authors reviewed the manuscript.
%\noindent
Competing Interests: The authors declare no competing interests.

\end{document}